\documentstyle[aps,twocolumn,psfig]{revtex}

\begin{document}

\draft

\wideabs{

\title{Charge and spin ordering in Nd$_{1/3}$Sr$_{2/3}$FeO$_{3}$}

\author{R. Kajimoto$^{\rm a}$\thanks{Corresponding
author. Fax: +81-29-283-3922. {\it E-mail address:}
kaji\@red.issp.u-tokyo.ac.jp (R. Kajimoto)}, Y. Oohara$^{\rm a}$,
M. Kubota$^{\rm a}$, H. Yoshizawa$^{\rm a}$, 
S. K. Park$^{\rm b}$, Y. Taguchi$^{\rm b}$, Y. Tokura$^{\rm b}$}

\address{$^{\rm a}$Neutron Scattering Laboratory, I. S. S. P.,
University of Tokyo, Tokai, Ibaraki, 319-1106, Japan \\
$^{\rm b}$Department of Applied Physics, University of
Tokyo, Bunkyo-ku, Tokyo 113-8656, Japan}

\date{\today}

\maketitle

\begin{abstract}
 We have investigated the charge and spin ordering in
 Nd$_{1/3}$Sr$_{2/3}$FeO$_{3}$ with neutron diffraction technique.  This
 sample undergoes a charge ordering transition accompanying charge
 disproportionation of $\mbox{2Fe}^{4+} \to \mbox{Fe}^{3+} +
 \mbox{Fe}^{5+}$. We measured the superlattice reflections due to the
 charge and spin ordering, and confirmed that charges and spins order
 simultaneously at $T_{\rm CO} = 185$ K. The ordering pattern of charges
 and spins in this sample can be viewed as three dimensional stripe
 order, and is compared with two dimensional stripe order observed in
 other transition metal oxides.
\end{abstract}

\pacs{{\it keywords:} A. oxides, B. crystal growth, C. neutron
scattering, D. charge-density wave, magnetic structure}

}


Charge ordering is widely seen in hole doped transition metal oxides,
such as cuprates (e.g. La$_{2-x}$Sr$_{x}$CuO$_{4}$ \cite{tranquada_cu}),
nickelates (e.g. La$_{2-x}$Sr$_{x}$NiO$_{4+\delta}$ \cite{tranquada_ni})
and manganites (e.g. Pr$_{1/2}$Ca$_{1/2}$MnO$_{3}$ \cite{tomioka95} and
Nd$_{1/2}$Sr$_{1/2}$MnO$_{3}$ \cite{kuwahara95}). Because the charge
ordering transition occurs at the region near the insulator-metal
transition or superconducting transition, a number of works concentrated
on the study of this phenomenon in order to clarify a key to understand
the origin of the insulator-metal transition or superconducting
transition. From these studies, it is widely recognized that the charge
ordering phenomenon is a consequence of the coupling or the competition
among the degrees of freedom of charge, spin, lattice, or orbitals

Hole doped perovskite-type $R_{1/3}$Sr$_{2/3}$FeO$_{3}$ is one of the
system which undergoes a charge ordering transition. In this system, the
charge ordering accompanies charge disproportionation of
$\mbox{2Fe}^{4+} \to \mbox{Fe}^{3+} + \mbox{Fe}^{5+}$ and simultaneous
antiferromagnetic spin ordering. A pioneering work with M\"{o}ssbauer
spectroscopy on La$_{1/3}$Sr$_{2/3}$Fe$_{3}$ by Takano {\it et al.}
revealed that there are two kinds of Fe ions with the ratio of $2:1$,
and they were attributed to Fe$^{3+}$ and Fe$^{5+}$
\cite{takano83}. This charge disproportionation state was confirmed by
Battle {\it et al.} with the neutron powder diffraction measurements on
the same compound. They observed the antiferromagnetic spin ordering
with sixfold periodicity along the cubic [111] direction and showed that
this magnetic structure was generated from the ordering of the layers of
the Fe$^{3+}$ ions and the Fe$^{5+}$ ions along the cubic [111]
direction in a sequence of
$\cdots\mbox{Fe}^{3+}\mbox{Fe}^{3+}\mbox{Fe}^{5+}\cdots$.  However, they
could not observe the superlattice reflections due to the charge
ordering presumably because the weak intensity of the reflections in the
powder sample, although it is well known that in many charge ordered
systems the ordering of charges strongly couples with lattice and
produces periodic modulation in the crystal structure
\cite{tranquada_cu,tranquada_ni,yoshi00,tomioka95,kuwahara95}. Recently
Li {\it et al.} observed the superlattice reflections of the charge
ordering in La$_{1-x}$Sr$_{x}$FeO$_{3}$ single crystals for the first
time by electron diffraction measurements \cite{li97}. Park {\it et al.}
showed that similar charge and spin ordering also exists in
$R_{1/3}$Sr$_{2/3}$FeO$_{3}$ single crystals where $R$ is rare-earth
atoms other than La \cite{park99}.

Neutron diffraction is very useful to investigated the spin and charge
coupled physics such as charge ordering, because it can detect the
direct evidence of the magnetic ordering and the structural modulations
related to the ordering of charges which is formed in a bulk
sample. Therefore we performed the neutron diffraction experiments on
one of the 2/3-hole-doped Fe oxides, Nd$_{1/3}$Sr$_{2/3}$FeO$_{3}$.  By
utilizing a single crystal sample, we could find the superlattice
reflections due to the structural modulations by the charge ordering
together with magnetic reflections. We could also detect a subtle change
of the nature of the charge ordering as a function of temperature.


A single crystal sample studied in the present study was grown by the
floating zone method in oxygen atmosphere with a traveling speed of 1.0
mm/h.  The detailed procedures of the sample preparation have already
been described elsewhere \cite{park99}. The quality of the sample was
checked by x-ray powder diffraction measurements and by electron probe
microanalysis.

Neutron diffraction experiments were performed using triple axis
spectrometer GPTAS installed at the JRR-3M reactor in JAERI, Tokai,
Japan with fixed incident neutron momentums $k_{\rm i} = 2.66$
\AA$^{-1}$ and 3.83 \AA$^{-1}$. The combination of collimators were
20$^{\prime}$-40$^{\prime}$-20$^{\prime}$-open and
40$^{\prime}$-80$^{\prime}$-80$^{\prime}$-80$^{\prime}$ (from
monochromator to detector).  Although the crystal structure of the
sample has a slight rhombohedral distortion along the cubic [111]
direction ($\alpha=60.1^{\circ}$ at 300 K), we employed the cubic
lattice ($a \sim 3.85$ \AA) notation of the scattering plane. The
sample was mounted in an Al can filled with He gas, and was attached to
the cold head of a closed-cycle helium gas refrigerator. The temperature
was controlled within an accuracy of 0.2 K.

\begin{figure}
 \centering \leavevmode
 \psfig{file=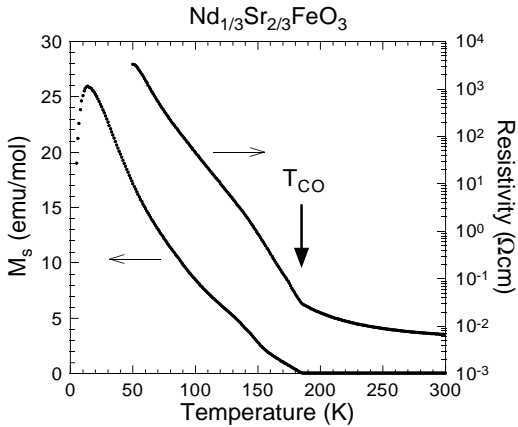,width=0.8\hsize}
 \vspace{2mm}
 \caption{Temperature dependence of resistivity and spontaneous
 magnetization for Nd$_{1/3}$Sr$_{2/3}$FeO$_{3}$.}
 \label{transport}
\end{figure}

First we show the results of resistivity and magnetization measurements
which were performed on the same crystal used in the neutron diffraction
study\cite{taguchi_un}. Figure \ref{transport} shows the temperature
dependence of the resistivity and the spontaneous magnetization. The
resistivity at room temperature is relatively low and gradually
increases as temperature decreases. However, it shows a steep increase
below $T_{\rm CO} = 185$ K because of the charge ordering transition. As
shown in Fig. \ref{transport}, small spontaneous magnetization appears
below $T_{\rm CO}$, signaling the onset of the antiferromagnetic
ordering with minute spin canting.

\begin{figure}
 \centering \leavevmode
 \psfig{file=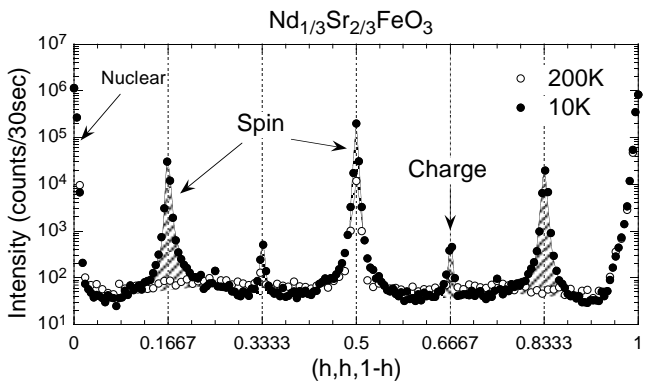,width=\hsize}
 \vspace{-1mm}
 \caption{A profile of the line scan along the $[11\bar{1}]$ direction
 at 200 K and 10 K. Shaded peaks represent the superlattice reflections
 due to the magnetic and charge ordering.}
 \label{profile}
\end{figure}

In order to characterize the charge and spin ordering, we surveyed the
$(hhl)$ scattering plane and found some superlattice reflections below
$T_{\rm CO}$. For an example of such survey scans, we show in
Fig. \ref{profile} profiles of line scans along $[11\bar{1}]$ direction
measured at 200 K and 10 K. One can see that at 10 K superlattice
reflections appear at $(\frac{1}{6},\frac{1}{6},\frac{5}{6})$,
$(\frac{1}{3},\frac{1}{3},\frac{2}{3})$,
$(\frac{1}{2},\frac{1}{2},\frac{1}{2})$,
$(\frac{2}{3},\frac{2}{3},\frac{1}{3})$, and
$(\frac{5}{6},\frac{5}{6},\frac{1}{6})$. The observed superlattice
reflections can be classified into three groups according to their
modulation vectors {\boldmath $q$}: $\mbox{\boldmath $q$}_{\frac{1}{6}}
= a^{*}(\frac{1}{6},\,\frac{1}{6},\,\frac{1}{6})$, $\mbox{\boldmath
$q$}_{\frac{1}{3}} = a^{*}(\frac{1}{3},\,\frac{1}{3},\,\frac{1}{3})$,
and $\mbox{\boldmath $q$}_{\frac{1}{2}} =
a^{*}(\frac{1}{2},\,\frac{1}{2},\,\frac{1}{2})$.  Due to the twin
domains of the cubic structure, we also observed reflections with
$\mbox{\boldmath $q$} = a^{*}(\frac{1}{6},\,\frac{1}{6},\,-\frac{1}{6})$
etc. in the $(hhl)$ zone.

$\mbox{\boldmath $q$}_{\frac{1}{3}}$ reflections were observed also by
electron diffraction measurements \cite{li97,park99} and they indicate
lattice modulation along the cubic [111] direction whose period was
three times larger than the lattice spacing of the (111) plane. This
modulation should originate from the charge ordering of
$\cdots\mbox{Fe}^{3+}\mbox{Fe}^{3+}\mbox{Fe}^{5+}\cdots$ and
accompanying local lattice distortion. The intensity of $\mbox{\boldmath
$q$}_{\frac{1}{6}}$ and $\mbox{\boldmath $q$}_{\frac{1}{2}}$ reflections
obey the $Q$ dependence of the magnetic form factor of the Fe
ion. Therefore they can be attributed to the antiferromagnetic ordering,
which is consistent with the previous powder neutron diffraction
measurements on La$_{1/3}$Sr$_{2/3}$FeO$_{3}$ \cite{battle90}. Note that
nuclear reflections were observed at
$(\frac{h}{2},\frac{h}{2},\frac{l}{2})$ even at $T > T_{\rm CO}$
(Fig. \ref{profile}). These reflections are forbidden in the previously
proposed rhombohedral symmetry $R\bar{3}c$, and indicate that the true
crystal symmetry of the present sample is lower than $R\bar{3}c$
\cite{battle90,li97,park99}.

The existence of $\mbox{\boldmath $q$}_{\frac{1}{6}}$ and
$\mbox{\boldmath $q$}_{\frac{1}{2}}$ magnetic modulation vectors means
that the distribution of magnetic moments can be represented by the sum
of two Fourier components, each has a wave vector of $\mbox{\boldmath
$q$}_{\frac{1}{6}}$ and $\mbox{\boldmath $q$}_{\frac{1}{2}}$. As a
consequence, if one assume the magnetic moments are localized on the Fe
sites, there are two Fe sites in the ratio of $2:1$ in the sequence of
$\cdots \Uparrow\, \Downarrow\, \downarrow\, \Downarrow\, \Uparrow\,
\uparrow\, \Uparrow \cdots$. The sites denoted by $\Uparrow$ and those
by $\uparrow$ may be attributed to Fe$^{3+}$ sites and Fe$^{5+}$ sites,
respectively.

In order to analyze the magnetic moments for two Fe sites, we have
performed neutron powder diffraction measurements to avoid difficulty in
analyzing the single crystal caused by the domain distribution. The
measurement was performed at 50 K, because below that temperature, the
reflections by the ordering of the magnetic moments of Nd$^{3+}$ ions
superpose the $\mbox{\boldmath $q$}_{\frac{1}{6}}$ magnetic reflections
(see below). The obtained value of the magnetic moments are 3.7
$\mu_{\rm B}$ for Fe$^{3+}$ site and 2.3 $\mu_{\rm B}$ for Fe$^{5+}$
site assuming the spins lie in the (111) plane. The direction of the
spins in the (111) plane could not be determined due to the high
symmetry. These values are similar to those obtained for
La$_{1/3}$Sr$_{2/3}$FeO$_{3}$ (3.61 $\mu_{\rm B}$ for Fe$^{3+}$ site and
2.72 $\mu_{\rm B}$ for Fe$^{5+}$ site) \cite{battle90}. The smaller
observed magnetic moments than their nominal values of 5 $\mu_{\rm B}$
and 3 $\mu_{\rm B}$ indicate the strong hybridization of the Fe 3d
orbitals and the O 2d orbitals because of the small charge-transfer
energy of Fe oxides \cite{bocquet92,matsuno99}.

\begin{figure}
 \centering \leavevmode
 \psfig{file=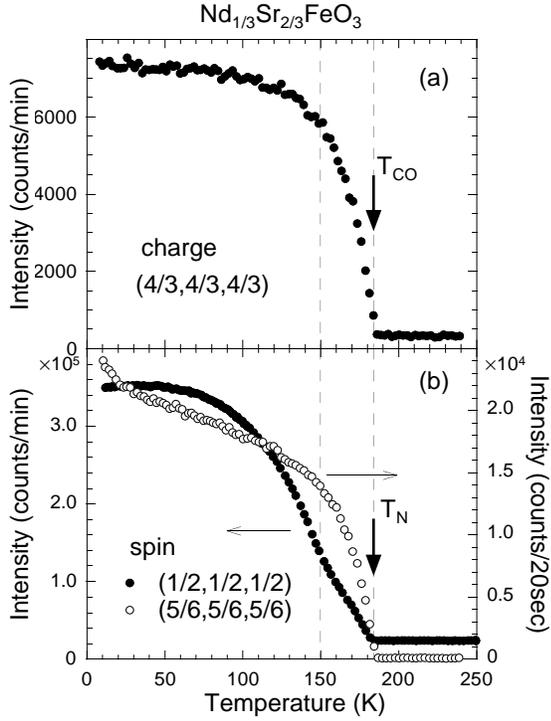,width=0.85\hsize}
 \vspace{2mm}
 \caption{Temperature dependence of the intensity of the spin and charge
 super lattice reflections. (a) Charge superlattice reflection
 $(\frac{4}{3}\frac{4}{3}\frac{4}{3})$. (b) Spin superlattice reflection
 $(\frac{5}{6}\frac{5}{6}\frac{5}{6})$ (open circles) and
 $(\frac{1}{2}\frac{1}{2}\frac{1}{2})$ (closed circles).}
 \label{Tdep}
\end{figure}

Figure \ref{Tdep} shows temperature dependence of the intensity of (a)
the $\mbox{\boldmath $q$}_{\frac{1}{3}}$ charge superlattice reflection,
(b) $\mbox{\boldmath $q$}_{\frac{1}{6}}$ (open circles) and
$\mbox{\boldmath $q$}_{\frac{1}{2}}$ (closed circles) spin superlattice
reflections. As decreasing temperature, all the reflection start to
develop at the same temperature $T_{\rm CO} = 185$ K, indicating that
charges and spins order simultaneously, which is consistent with the
resistivity and magnetization data shown in Fig. \ref{transport}. The
transition at $T_{\rm CO}$ is first order with small hysteresis of $\sim
4$ K. The increase of the intensity of the
$(\frac{5}{6}\frac{5}{6}\frac{5}{6})$ reflection below $T \sim 50$ K
comes from the ordering of the spins of Nd$^{3+}$ ions.

The development of the charge ordering almost saturates around $T_{\rm
s} = 150$ K. The character of the spin ordering also changes at this
temperature. The slope of the curve for the $\mbox{\boldmath
$q$}_{\frac{1}{2}}$ reflection becomes steeper below $T_{\rm s}$, while
the growth of the $\mbox{\boldmath $q$}_{\frac{1}{6}}$ reflection is
slightly suppressed around $T_{\rm s}$.

The anomaly in the spin superlattice reflections at $T_{\rm s}$ can not
be explained by the rotation of the spin orientations. Because the
scattering vectors of the two kinds of the spin reflections shown in
Fig. \ref{Tdep} (b) are parallel, the rotation of the spins should
produce the same effect on the intensity of both spin
reflections. Therefore, the anomaly at $T_{\rm s}$ should be attributed
to the change of the ratio of the two Fourier component of the spin
density wave at this temperature. If one assume the moments are
localized at Fe sites, this means that the moments on Fe$^{3+}$ sites
increase while those on Fe$^{5+}$ sites decrease, suggesting the
enhancement of the charge disproportionation. Of course this
interpretation is too na\"{\i}ve because the holes may locate on oxygen
sites and the magnetic moments distribute continuously from site to site
due to the hybridization of the Fe 3d orbitals and the O 2p
orbitals. Nevertheless, from the fact that the charge ordering almost
saturates around $T_{\rm s}$, one can say that the change of the
distribution of the magnetic moments are correlated with the charge
ordering, and the nature of the charge and the spin ordering changes
when the charge ordering sufficiently develops.  We should note that we
also observed the similar anomaly at $T < T_{\rm CO}$ in another
2/3-hole-doped charge disproportionated Fe oxides
Pr$_{1/3}$Sr$_{2/3}$FeO$_{3}$ \cite{oohara_un}.

One of the most interesting phenomena in the charge ordering transition
in the hole-doped transition metal oxides is stripe order, where the
doped holes align to form domain walls and spins order
antiferromagnetically. Most of the stripe ordering observed so far, e.g.
in La$_{2-x}$Sr$_{x}$CuO$_{4}$ \cite{tranquada_cu} or
La$_{2-x}$Sr$_{x}$NiO$_{4+\delta}$ \cite{tranquada_ni,yoshi00}, is two
dimensional (2 D) which has one dimensional domain walls. On the other
hand, the charge and spin ordering pattern of the 2/3-hole-doped cubic
perovskite Fe oxides including Nd$_{1/3}$Sr$_{2/3}$FeO$_{3}$ can be
viewed as a three dimensional (3 D) stripe order propagating the [111]
direction with 2 D domain walls parallel to (111) planes,

One of the clear differences between the 2 D and the 3 D stripe order is
the relation between the charge ordering temperature $T_{\rm CO}$ and
the spin ordering temperature $T_{\rm N}$. In the 2 D stripe order,
spins always order {\em after} the charges order, while in the 3 D
stripe order, spins and charges order {\em simultaneously}. For the
formation of the stripe order, the importance of the superexchange
interaction between the spins as well as the Coulomb interaction between
charges is widely recognized, although there is still a dispute about
the driving force for the stripe order.  Zachar {\it et al.} proposed a
phase diagram for the stripe order based on Landau
theory\cite{zachar98}. They claimed that there are several ways in the
transition to the stripe ordered state: a charge driven transition, a
charge-spin coupling driven transition, and a spin driven transition. In
the charge driven transition, $T_{\rm N}$ should be lower than $T_{\rm
CO}$, while in the charge-spin coupled transition and in the spin driven
transition $T_{\rm N}$ should be same as $T_{\rm CO}$. Moreover, the
spin-charge coupled transition should be the first order while the spin
driven transition should be the second order. In their framework, the 2
D stripe order observed in cuprates or nickelates is classified into the
charge driven transition, and the 3 D stripe order observed in the
present study can be classified into the spin-charge coupling driven
transition.  In any event, the fact that spins and charges order
simultaneously in the 3 D stripe order means that the role of the spin
ordering in the stripe order becomes relatively important compared to
the 2 D stripe.

We think the increase of the relative importance of the spin ordering in
the 3 D stripe can be interpreted as a consequence of the difference in
the dimensionality between the 2 D and the 3 D stripe. In the 2 D stripe
order, the number of the nearest holes around a hole in a domain wall
becomes larger because the domain walls become 2 D. Therefore, the
energy loss by the Coulomb repulsion between the holes may become larger
as compared to the 3 D stripe order. On the other hand, the number of
the nearest sites for the undoped region also becomes larger because of
the three dimensionality, which may increase the energy gain by the
superexchange interaction between spins.

In order to verify the above scenario, the energy scale of the Coulomb
interaction and the superexchange interaction should be examined. We
would like to note that by recent Hartree-Fock calculations, it has been
shown that the 3 D stripe ordering observed in Fe oxides can be well
explained by the superexchange interaction \cite{mizokawa98}, which
indicates the importance of the spin interaction for the formation of
the stripe order.


In summary, we have investigated the charge and spin ordering in a
Nd$_{1/3}$Sr$_{2/3}$MnO$_{3}$ crystal using neutron diffraction
technique. We measured the superlattice reflections due to the charge
and spin ordering, and confirmed that charges and spins order
simultaneously at $T_{\rm CO} = 185$ K. The character of the charge and
spin ordering changes at 150 K when the charge ordering almost
saturates. The pattern of the charge and spin ordering in this sample
can be viewed as the 3 D stripe order, and the transition may be driven
by the spin-charge coupling.


This work was supported by a Grant-In-Aid for Scientific Research from
the Ministry of Education, Science, Sports and Culture, Japan and by the
New Energy and Industrial Technology Development Organization (NEDO) of
Japan.

\end{document}